\documentclass[11pt,preprint]{aastex}

\shorttitle{Cluster Surrounding V838 Mon}
\shortauthors{Af\c{s}ar \& Bond}

\begin{document}

\title{A Young Stellar Cluster Surrounding the Peculiar Eruptive Variable
V838~Monocerotis\altaffilmark{1}}

\author{Melike Af\c{s}ar\altaffilmark{2} and Howard E. Bond}

\affil{Space Telescope Science Institute, 3700 San Martin Drive, Baltimore, MD
21218; melike.afsar@ege.edu.tr, bond@stsci.edu}

\altaffiltext{1}
{Based on observations made with the Small- and
Medium-Aperture Research Telescope System (SMARTS).}

\altaffiltext{2}
{Current address: Department of Astronomy and Space Sciences, Ege University,
35100 Bornova, \.{I}zmir, Turkey}



\begin{abstract}

V838~Monocerotis is an unusual variable star that underwent a
sudden outburst in 2002. Unlike a classical nova, which
quickly evolves to high temperatures, V838~Mon remained an
extremely cool, luminous supergiant throughout its eruption.
It continues to illuminate a spectacular series of light
echoes, as the outburst light is scattered from nearby
circumstellar dust. V838~Mon has an unresolved B3~V companion
star.

During a program of spectroscopic monitoring of V838~Mon, we
serendipitously discovered that a neighboring 16th-mag star is
also of type~B\null. We then carried out a spectroscopic
survey of other stars in the vicinity, revealing two more
B-type stars, all within $45''$ of V838~Mon. We have
determined the distance to this sparse, young cluster, based
on spectral classification and photometric main-sequence
fitting of the three B~stars. The cluster distance is found to
be $6.2\pm1.2$~kpc, in excellent agreement with the geometric
distance to V838~Mon of 5.9~kpc obtained from {\it Hubble
Space Telescope\/} polarimetry of the light echoes. An upper
limit to the age of the cluster is about 25~Myr, and its
reddening is $E(B-V)=0.85$.

The absolute luminosity of V838~Mon during its outburst, based
on our distance measurement, was very similar to that of
M31~RV, an object in the bulge of M31 that was also a cool
supergiant throughout its eruption in 1988. However, there is
no young population at the site of M31~RV.

Using our distance determination, we show that the B3~V
companion of V838~Mon is sufficient to account for the entire
luminosity of the variable star measured on sky-survey
photographs before its outburst.  The B3 star is currently,
however, about 1~mag fainter than before the eruption,
suggesting that it is now suffering extinction due to dust
ejected from V838~Mon.  These results indicate that, whatever
the nature of the progenitor object, it was not of high
luminosity. Nor does it appear possible to form a nova-like
cataclysmic binary system within the young age of the V838~Mon
cluster. These considerations appear to leave
stellar-collision or -merger scenarios as one of the remaining
viable explanations for the outbursts of V838~Mon and M31~RV.

\end{abstract}



\section{Introduction}


The outburst of the previously unknown variable star V838 Monocerotis was
discovered in 2002 January by Brown (2002).  By early February V838~Mon reached
6th magnitude, but by 2002 May it had returned to quiescence at optical
wavelengths. Shortly after maximum light, an expanding light echo was discovered
by Henden, Munari, \& Schwartz (2002). These light echoes have evolved to
become  the most spectacular display of the phenomenon in astronomical history.
They have been the subject of extensive imaging by ground-based observers (e.g.,
Crause et al.\  2005 and references therein) and with the {\it Hubble Space
Telescope\/} ({\it HST\/}) (Bond et al.\ 2003, 2006).

The eruption of V838 Mon was of a very unusual type. In a classical-nova
outburst, the ejecta expand rapidly, become optically thin, and expose an
extremely hot source.  By contrast, V838~Mon remained extremely cool throughout
its outburst, becoming one of the coolest known stars---in fact it has been
called the first L-type supergiant (Evans et al.\ 2003). In 2005 it developed
rapidly strengthening SiO maser emission (Claussen et al.\ 2005). A variety of
explanations for the outburst have been proposed, many of them mutually
exclusive (see the recent summary in Tylenda \&  Soker 2006, and the forthcoming
proceedings of an international conference on V838~Mon---Corradi \& Munari
2006). These explanations involve either thermonuclear processes (an unusual
nova-like outburst on a white dwarf, a thermonuclear event in a massive star, or
a helium shell flash in a post-AGB star), or the release of gravitational energy
(through stellar or planetary mergers or collisions).

A possible new constraint on the nature of V838~Mon came from the spectroscopic
discovery of a B3~V companion to the variable star (Munari \& Desidera 2002;
Wagner \& Starrfield 2002). The companion is unresolved even in {\it HST\/}
images.\footnote{In an unpublished analysis, we have examined the {\it HST\/}
images of V838~Mon obtained in late 2002 (which have been discussed by Bond et
al.\ 2003). At that time the $B$ light was dominated by the B3~V star, while the
cool outbursting star dominated at $V$ and $I$\null. We detect no centroid shift
of greater than $0\farcs1$ between the blue and red images.} However, it is not
close enough to the variable to have been engulfed when V838~Mon expanded to a
radius of several AU.

Two other eruptive objects have attracted attention as possible analogs of
V838~Mon. One of them is the ``M31 red variable," or ``M31 RV,'' which underwent
an outburst in mid-1988 that was remarkably similar to that of V838~Mon (Bond \&
Siegel 2006 and references therein), although not as well observed.  M31~RV
occurred in the nuclear bulge of the Andromeda Galaxy, and is thus at a known
distance. The other is V4332~Sgr, a Galactic object that had a nova-like
outburst in 1994 during which it likewise remained very cool (Martini et al.\
1999; Tylenda et al.\ 2005 and references therein). 

In this paper, we present our serendipitous discovery that V838~Mon is a member
of a small, young open cluster. We will use spectral classification and
photometry of the cluster members to determine a distance, which we will compare
with the direct geometric distance to V838~Mon derived from {\it HST\/}
polarimetry of the light echoes. We will also derive a limit to the cluster's
age, and will compare the stellar populations surrounding V838~Mon and
M31~RV\null. We will close with a brief discussion of the V838~Mon progenitor
object, its B3 companion, and some new constraints on the outburst mechanism
that result from our observations.

\section{Observations and Data Reduction}

\subsection{Spectroscopy}

We have been monitoring the spectroscopic development of V838~Mon since early
2003, using the SMARTS 1.5-m telescope located at Cerro Tololo Interamerican
Observatory (CTIO) and the Boller \& Chivens CCD spectrograph. Most of our
observations have been obtained with a 600 groove~mm$^{-1}$ grating used in
first order (grating designation ``26/I''), yielding a FWHM resolution of
4.3~\AA\ and a wavelength coverage of 3530--5300~\AA\null.  The CCD images are
bias-subtracted and flat-fielded using standard IRAF\footnote{IRAF is
distributed by the National Optical Astronomy Observatory, which is operated by
the Association of Universities for Research in Astronomy, Inc., under
cooperative agreement with the National Science Foundation.} routines, and the
spectra are then extracted from the images using the {\it apall\/} task.
Wavelength calibration is accomplished using He-Ar comparison-lamp exposures
taken before and after each stellar exposure.

Our data are long-slit spectra, in which the slit length projected onto the sky
is about $6'$, in an east-west direction. It is not unusual for neighboring
field stars to fall onto the slit, but we were surprised when a 16th-mag star
lying on the slit almost directly east of V838~Mon proved, entirely
serendipitously, to have a B-type spectrum.  Although V838~Mon lies at a low
galactic latitude ($\ell=217\fdg8$, $b=+1\fdg0$), a B-type star as faint as
16th~mag would lie at the outskirts of the Galactic disk. This makes its
existence very unusual, especially when lying within a few arcseconds of a star
that is itself extraordinary. Adding this to the fact that V838~Mon has
an unresolved B~companion made it appear very likely that the serendipitous star
is at the same distance. This in turn raised the possibility that there could be
further faint early-type objects in the field surrounding V838~Mon.

To investigate this possibility, we have used the SMARTS 1.5-m telescope to
obtain spectra of several more stars in the immediate vicinity of V838~Mon. For
these exploratory observations we used a 150 groove~mm$^{-1}$ grating
(designation ``13/I''), giving a resolution of 17.2~\AA\ and coverage
3150--9375~\AA\null. Most of the neighboring stars have proven to be unrelated
foreground stars, but our observations to date have disclosed two further
14th-15th mag B-type stars near V838~Mon. All three of these early-type stars
lie within $45''$ of the variable, or within a projected separation of only
1.3~pc if the distance is $\sim$6~kpc (see below). Thus there is little doubt
that V838~Mon is accompanied by a previously unrecognized sparse, young cluster.

Table~1 lists some properties of the new B-type stars near V838~Mon. The
coordinates are taken from the NOMAD-1.0 astrometric catalog  (Zacharias et al.\
2004). Other information in Table~1 is explained below.

Figure~1 presents a finding chart that identifies the new B~stars, using an
arbitrary numbering scheme. Star~7 is the serendipitous B star discovered first,
lying nearly directly east of V838~Mon; the other two B stars are numbers 8
and~9. All of the other numbered stars marked in Figure~1 are unrelated
foreground stars, based on our 1.5-m spectroscopic observations. (In addition,
the bright star $\sim$$20''$ to the northeast of star~4 has colors showing it to
belong to the foreground, based on our photometry described below.) Figure~1
illustrates that the cluster is not obvious in a direct image, and as we will
see below the reddened B-type members and foreground F-G stars have similar
colors; hence spectroscopic observations are the only practical means for
identifying cluster members in this crowded, low-latitude field.

As this paper was being prepared, we became aware that Wisniewski, Bjorkman, \&
Magalh{\~a}es (2003; hereafter WBM2003) had already pointed out that our three
stars are likely to lie at a similar distance to V838~Mon itself, based on the
similar polarimetric properties of all four stars. Wisniewski et al.\ used a
different numbering scheme for these stars, and we give cross references to
their star numbers in the second column of Table~1.

Photometry for the field surrounding V838~Mon has been published by Munari
et~al.\ (2005; hereafter M05), and cross references to their numbering system
are given in the third column of Table~1.

\subsection{Spectral Classification}

Following the serendipitous discovery of the first field B-type star, and our
subsequent discoveries of two more B stars with the low-resolution 13/I grating
setup, we obtained moderate-resolution of the latter two with the SMARTS 1.5-m
spectrograph and the moderate-resolution 26/I grating. In Figure~1 we show these
spectra, along with spectra of several classification standards taken with the
same setup. The spectra of stars~8 and~9 are exposures of $3\times800$~s and
$3\times600$~s, respectively, obtained in 2004 December, while the spectrum of
star~7 is derived from 26/I spectra of V838~Mon taken on 20 nights in 2003-2004
in which star~7 also lies on the slit. The exposures on V838~Mon were
typically $3\times900$~s on each night, but since star~7 is not precisely east
of the variable not all of its light falls within the slit. Also included in
Figure~1 is a spectrum of V838~Mon itself based on observations on five nights
between 2003 February and May, a time when the variable had declined
considerably from its 2002 outburst, but was still contributing some light even
in this blue spectral range (note the TiO bands longward of H$\beta$, for
example).

We classified the three B stars based both upon a visual comparison with the
standard stars shown in Figure~2, and upon equivalent-width measurements of the
\ion{He}{1} and Balmer lines. The results are given in Table~1 and are believed
to be accurate to better than one spectral subtype. We were not able to 
classify the unresolved B-type companion of V838~Mon itself with these methods
because of the contamination from the cool component. However, from a comparison
of the equivalent widths of the bluest Balmer lines (H8, H9, and H10) in
V838~Mon (where the contamination is lowest) and in the standard stars, we find
a type of B3~V\null. This agrees well with the other authors cited in \S1.

%
%

\subsection{Photometry}

We have also been monitoring the field of V838~Mon regularly since early 2003
with the SMARTS 1.3-m telescope and the ANDICAM optical/near-infrared direct
camera (DePoy et~al.\ 2003).  In order to derive calibrated photometry of the
stars surrounding V838~Mon, we selected frames obtained on five photometric
nights in 2003-2004 on which the Ru~149 standard field from Landolt (1992) had
been observed at a similar airmass (so as to essentially remove the dependence
of the calibration upon variations in atmospheric extinction coefficients). 

We used standard IRAF routines for bias subtraction and flat-fielding of the
2003 CCD frames (starting in 2003 August, however, ANDICAM data have been
reduced through a pipeline at Yale University before distribution to observers).
We then used the IRAF implementation of DAOPHOT (Stetson 1987) for star finding
and to obtain point-spread-function (PSF) instrumental magnitudes for the stars.
These magnitudes were then transformed to the Johnson $BV$ system using the
Landolt standards. The results are given in Table~2. The internal errors in $V$
and $B-V$ are about $\pm$0.004 and  $\pm$0.005~mag, respectively, for the two
brighter stars, and about $\pm$0.005 and  $\pm$0.007~mag, respectively, for
star~7. However, since typically only one Landolt field was observed per night,
the external errors are probably closer to $\pm$0.01-0.02~mag.

When we compared our photometry with that published by M05 for the same three
stars, we found very large discrepancies (up to $\sim$0.7~mag in $V$), and we
also noted that unusually large errors are listed by M05 for two of the three
stars. Variability does not account for the large discrepancies, since we find
constancy for all three stars, leaving us with no clear
explanation.\footnote{Bacher et~al.\ (2005, caption to their Fig.~2) note that
the M05 photometry was obtained while V838~Mon was very bright, thus providing a
possible explanation for the poor photometry for stars very close to the
variable. As this paper was being completed, we received a private communication
from A.~Henden indicating that his recent update of the M05 photometry shows
much better agreement with our results.} We have performed several checks on our
photometric measurements, and are confident that our results are reliable.

\section{Distance and Other Properties of the Cluster}

\subsection{Spectroscopic Parallax}

We can now calculate distances to each of the three B stars, and thus determine
a distance to the cluster. Details of the calculations are given in Table~2. The
first four columns repeat the star designations, spectral types, and photometry
from Table~1. The fifth column gives the intrinsic color corresponding to each
spectral type, $(B-V)_0$, taken from the tabulation of Schmidt-Kaler (1982). The
sixth column gives the estimated color excess, $E(B-V) = (B-V)-(B-V)_0$. The
color excesses agree well among the three stars, with a mean of $E(B-V) = 0.84$
and a scatter of about $\pm$0.02~mag. This determination agrees well with those
of other authors. For example M05 found $E(B-V) = 0.87\pm0.01$ using several
different methods. Tylenda (2005) discusses recent reddening determinations by
several authors, and adopts $E(B-V) \simeq 0.9$. It should be noted, however,
that these authors have given high weight to methods based on the unresolved B3
companion of V838~Mon, which (as discussed below) may suffer some local
extinction in addition to foreground interstellar extinction.

The next column in Table~2 gives the magnitude of each star corrected for
extinction, $V_0$, calculated assuming $A_V=3.1E(B-V)$\null. Column~8 gives the
absolute magnitude corresponding to each spectral type, $M_V$, again taken from
Schmidt-Kaler (1982). The final column gives the distance moduli, $(m-M)_0$,
whose mean is 14.0, or a distance of 6.3~kpc. 

The cluster distance modulus has an internal error of about $\pm$0.2~mag, based
on the scatter among the three stars. However, systematic errors are undoubtedly
larger. This is indicated by the scatter among different calibrations of the
relation between spectral types and absolute magnitudes (e.g., Lesh 1968;
Schmidt-Kaler 1982; Cramer 1997), which amounts typically to about $\pm$0.4~mag.
At a distance of 6.3~kpc, this corresponds to an error of $\pm$1.2~kpc.

%

\subsection{Main-sequence Fitting}


We can also estimate the distance to our cluster through photometric
main-sequence fitting. In the absence of any spectroscopic information, this
method would suffer from the well-known near-degeneracy between extinction and
distance for early-type stars; this is due to fact that the main sequence lies
along a steep, nearly straight line in the $V,(B-V)$ diagram. However, with the
additional constraints from the spectral types, main-sequence fitting become
possible.

To define the main sequence, we chose the lightly reddened open cluster
NGC~2362, which, at an age of $\sim$5~Myr, has been described as a template for
early stellar evolution (Moitinho et~al.\ 2001; hereafter MAHL01). This
cluster's unevolved main sequence extends to type B1~V (Johnson \& Morgan 1953).
We took $B,V$ photometry for NGC~2362 from Johnson \& Morgan and from Perry
(1973), and adopted $E(B-V)=0.10$ and $d=1.48$~kpc from MAHL01. We then
corrected the photometry to the spectroscopic reddening and distance for the
V838~Mon cluster found above.  The match with the three V838~Mon B stars was
very good, so we applied just one iteration of adjusting first the reddening of
the B stars and then the distance, so as to improve the fit. This resulted in a
V838~Mon cluster reddening of $E(B-V)=0.85$ and distance modulus $(m-M)_0=13.97$
($d=6.2$~kpc).  The external errors here are similar to those for the
spectroscopic parallax, since MAHL01 based the distance to NGC~2362 on the
Schmidt-Kaler (1982) zero-age main sequence.

In Figure~3 we plot the $V,(B-V)$ values for the three B stars as large filled
blue circles.  The small open red circles are the Johnson-Morgan and Perry
photometry of NGC~2362, adjusted to the V838~Mon reddening and distance found in
the previous paragraph. The fit is excellent, and strongly supports the
conclusion that our three B stars do form a physical cluster.

A direct geometrical distance determination for V838~Mon has been carried out by
Sparks et al.\ (2006), based on polarimetric imaging of the light echo obtained
with {\it HST\/}\null. This result, 5.9~kpc, is in excellent agreement with our
determination of 6.2~kpc based on the associated B-type stars.

Also plotted in Figure~3, as small filled black circles, is our photometry for
all stars within a radius of $90''$ of V838~Mon. For most of these stars we do
not know whether they are cluster members or not, but we do have SMARTS 1.5-m
spectra for six of them that establish them as belonging to the foreground;
these non-members are marked with black stars in Figure~3, and they are numbered
in the finding chart in Figure~1. 

As Figure~3 shows, the foreground F- and G-type stars have similar colors to the
reddened B-type cluster members. This demonstrates that spectroscopic
observations are essential to the identification of cluster members. It would be
very interesting to have spectra of the other stars in the vicinity of V838~Mon,
especially below $V\simeq17.5$, where the slope of the main sequence changes;
any of these candidates that proved to be cluster members would provide tighter
constraints on the cluster reddening and distance.


We can compare the luminosity of V838~Mon with that of the apparently similar
object M31~RV (see \S1). At maximum light (2002 February 6) V838~Mon had $B=7.9$
(M05, their Figure~1). For the reddening and distance derived here, this
corresponds to an absolute blue magnitude at maximum of $M_B=-9.6$. The
brightest $B$ magnitude measured for M31~RV during its 1988 outburst was 17.3
(Bryan \& Royer 1992; Boschi \& Munari 2004). The latter authors, adopting
$E(B-V)=0.12$ and $(m-M)_0=24.48$ for M31, found an absolute magnitude of
$M_B=-7.7$.  However, the light curve of M31~RV was very poorly sampled around
its maximum. It is more meaningful to compare the luminosities of the two
objects at the same well-covered portions of both light curves. Referring to
Figure~2 of Boschi \& Munari (2004), and using their light curves in
Kron-Cousins $R$, we can compare the luminosities at a point just before the
rapid fading in the $R$ band.  For M31~RV, this brightness is $R\simeq15$, and
for V838~Mon it is $R\simeq6.2$.  Correcting for reddening and distance, the
corresponding absolute magnitudes are $M_R\simeq-9.8$ and $-9.7$, respectively.
Thus, at least at this stage in their outbursts, the absolute luminosities were
nearly identical.\footnote{However, the referee has pointed out that there are
some spectroscopic dissimilarities between V838~Mon and M31~RV. For example, a
comparison of the spectrum of V838~Mon of Evans et al.\ (2003, their Fig.~2) and
of M31~RV by Rich et al.\ (1989, their Fig.~4) shows that the former had much
stronger TiO bands.}

\subsection{Diffuse Interstellar Bands}

%

As shown in Figure~2, the spectra of our three B-type stars exhibit conspicuous
diffuse interstellar band (DIB) features centered near 4428~\AA\null. We measure
the average equivalent width of the 4428~\AA\ DIB in these three stars to be
about 3.1~\AA\null. Snow, Zukowski, \& Massey (2002, their Fig.~6) show that
there is a correlation between the DIB equivalent width and the amount of
interstellar extinction in a large sample of Galactic stars, but with
considerable scatter. At a reddening of $E(B-V)=0.85$, our mean equivalent width
is at the upper envelope of the values plotted by Snow et~al.

These strong DIBs will provide a useful spectroscopic discriminant in future
attempts to identify faint members of the V838~Mon cluster.

\subsection{Cluster Age and Stellar Masses}

Since the three B stars all appear to lie on the zero-age main sequence, we can
only set an upper limit to the age of the V838~Mon cluster. By reference to the
isochrones and evolutionary tracks\footnote{Available at
http://www.te.astro.it/BASTI/index.php} of Pientrinfini et al.\ (2006), we find
that this upper limit is about 25~Myr. According to these isochrones, the masses
of stars~7, 8, and 9 are about 4.7, 6.4, and $6.7\,M_\odot$, respectively.


\section{Discussion}

\subsection{The Stellar Populations of V838 Mon and M31 RV}

V838~Mon is accompanied by a previously reported, unresolved B3~V companion. In
this paper we have shown that V838~Mon also belongs to a sparse cluster,
containing at least three other members, lying at the outer edge of the Milky
Way disk. The main sequence of this cluster extends up to spectral type B3~V,
implying an age of less than 25~Myr.

We could thus be tempted to speculate that the outburst of V838~Mon in 2002
represents an evolutionary event that occurs in stars of masses
$\ga$7--$8\,M_\odot$\null. We could even speculate that such events occur in
stars whose masses lie near the dividing line between those that explode as
Type~II supernovae and those that become white dwarfs more quiescently.

Unfortunately, such a speculation appears to be dashed by the recent study of
M31~RV by Bond \& Siegel (2006). They used archival {\it HST\/} images that
serendipitously included the outburst site of M31~RV, and showed that the
population surrounding the object contains exclusively old, low-mass stars
belonging to the nuclear bulge of M31. There is no young population at all at
this site, let alone bright B stars younger than 25~Myr.

Thus, if the outbursts of both V838~Mon and M31~RV arose from a common
mechanism, it is a mechanism that can occur among both very young and very old
stars.   

\subsection{The Underluminous B3~V Companion of V838 Mon}


Several authors (e.g., Tylenda, Soker, \& Szczerba 2005 and references therein) 
have attempted to determine the brightness of V838~Mon before its 2002 outburst.
The available material for such a determination at optical wavelengths comes
from the various photographic sky surveys, and from sky-patrol photographic
plates. Archival patrol plates show that the star did not vary over the interval
from 1928 to 1994 (Goranskij et al.\ 2004); however, Kimeswenger \& Eyres (2006)
have provided evidence suggesting that there was some fading of V838~Mon in the
late 1990's, shortly before the onset of the outburst.

In an analysis of high-resolution digitizations of a SERC survey plate obtained
in 1983 with a J emulsion, Kimeswenger \& Eyres (2006) report that V838~Mon had
a $B_J$ magnitude of $15.49\pm0.09$. Adopting a transformation equation of
$B=B_J+0.28(B-V)$ (Bacher, Kimeswenger, \& Teutsch 2005 and references therein),
and assuming that V838~Mon at that time had the same $B-V$ color as our star~9,
we find that its $B$ magnitude in 1983 was 15.66. (Goranskij et al.\ 2004, in an
independent analysis based on the lower-resolution scans of the plate material
available in the Digitized Sky Survey, find a slightly fainter pre-outburst
value of $B=15.81\pm0.06$.)

The companion of V838~Mon, if it is a normal B3~V star, would be expected to
have a luminosity similar to that of our star~9, also classified B3~V\null.
According to our photometry (Table~1), star~9 has a $B$ magnitude of 15.41. 
This is in fact slightly brighter than the actual pre-outburst $B$ magnitude of
V838~Mon found above, suggesting that {\it the progenitor of the outbursting
object contributed a negligible fraction of the pre-eruption light.} (Even if
the unresolved companion were in fact more similar to our slightly later star~8,
which we classified B4~V and has $B=15.63$, it still would have contributed {\it
all\/} of the pre-outburst light.)


It is well established that when V838~Mon subsided back to quiescence (at
optical wavelengths) in mid-2002, it was significantly fainter than before the
outburst (see, for example, the light curves in Figure~1 of M05, and those given
by Goranskij et al.\ 2004).  This
faintness has continued to the present time; for example, our own recent SMARTS
1.3-m observation, obtained on 2006 February 19, shows V838~Mon at $B=16.61$.
(Actually, this is an upper limit to the current brightness of the B3 star,
since there is some contribution of light from the outbursting companion;
however, our spectrum of V838~Mon in Figure~2 indicates that the contribution in
the $B$ band is relatively small.)

In order to explain this fading by $\sim$1~mag, several authors have argued that
the progenitor of the outbursting object contributed significant optical light
to the pre-outburst magnitude, and that this extra light is now absent or
diminished. M05, for example, concluded that one component of the binary was a
massive star that underwent a thermonuclear event in 2002. 

Our observations and distance/reddening determination, however, indicate that
the pre-outburst optical light can be attributed {\it entirely\/} to the B3
companion, and we know that this companion {\it is still present}. The actual
issue, then, is to explain why the B3 star is now about one magnitude fainter at
$B$ than it was before the eruption. We suggest that the explanation is rather
prosaic: the B3 companion is close enough to the outbursting component that it
is partially submerged in the dust that is being produced copiously and is
relatively slowly flowing away from V838~Mon.\footnote{Recent blue spectra of
V838~Mon show a strengthening of emission lines due to [\ion{Fe}{2}] (Barsukova
et al.\ 2006 as well as our own SMARTS spectra). Barsukova et al.\ attribute the
excitation of these lines to the B3 star, and suggest this indicates that
outburst ejecta have reached the vicinity of the B star.} (Alternatively, the B3
star could be {\it behind\/} the cool supergiant, with its light shining through
the dust envelope.)

If this is true, then we do not need to account for any extra light in the
system before the 2002 outburst.  This appears to rule out the presence of any
star as massive as star~9, either near the main sequence or in an evolved, even
more luminous state (as was proposed by M05). 

The B3 companion would also be expected to be reddened by a larger amount than
the cluster, if our scenario is correct. However, it is difficult to isolate the
color of the B3 star at the present time, because V838~Mon itself still
contributes significantly to the system light at the $V$ band and longward.

%

\subsection{Conclusion}

Our observations have revealed that V838~Mon belongs to a young cluster. This
discovery has, however, only deepened the enigma of V838~Mon and the similar
object M31~RV, because we now know that the former arose from a very young
population, whereas the latter belongs to a very old population.

Moreover, our discussion suggests that virtually all of the pre-outburst light
of the V838~Mon binary was due to the unresolved B3 companion.  

What, then, was the nature of the progenitor objects that produced the outbursts
of V838~Mon and M31~RV\null? They apparently can exist at both ends of the range
of stellar ages. The luminosity of the progenitor before outburst, at least in
the case of V838~Mon, was small compared to that of a B3 dwarf. That would be
consistent with the low luminosity of a typical nova-like cataclysmic variable.
However, the young age of the V838~Mon cluster would appear to rule out the
evolutionary timescale required to produce an accreting white dwarf in a compact
binary, a requirement of a nova-like outburst mechanism. 

It may be that the stellar-merger scenario advocated in a recent series of
papers (e.g., Tylenda \& Soker 2006 and references therein; see also Retter et
al.\ 2006, who advocate a planet-star merger) can satisfy these new constraints,
provided that collisions between low-mass stars (which exist in all populations)
could produce the required high luminosities, and can be shown to occur often
enough. At the moment, however, the nature of these extraordinary outbursts
remains one of the leading unsolved problems in stellar astrophysics.

\acknowledgments

We thank the excellent SMARTS 1.3- and 1.5-m service observers for obtaining the
data: Claudio Aguilera, Juan Espinoza, David Gonzalez, Sergio Gonzalez, and
Alberto Pasten. Fred Walter, Rebecca Winnick, and Jenica Nelan have carried out
the telescope scheduling, and data distribution is done by Suzanne Tourtellotte.
M.~A. would like to thank The Scientific and Technological Research Council of
Turkey (TUBITAK) and her advisor Prof.\ Cafer \.{I}bano\v{g}lu, as well as H. E.
B. and STScI for their support during her doctoral studies. STScI's
participation in the SMARTS consortium is supported by the STScI Director's
Discretionary Research Fund. We thank numerous colleagues for discussions of
V838~Mon, including especially Lisa Crause, Romano Corradi, Arne Henden, Stefan
Kimeswenger, Zolt Levay, Ulisse Munari, Nino Panagia, William Sparks, Sumner
Starrfield, Ben Sugerman, Mark Wagner, John Wisniewski, and all of the
participants at the La Palma conference on ``The Nature of V838 Mon and its
Light Echo.'' 

This research has made use of the WEBDA open-cluster database, operated at the
Institute for Astronomy of the University of Vienna.

H. E. B. dedicates this paper to the memory of his friend Charles L. Perry.

\clearpage


\begin{figure}
\begin{center}
\includegraphics[width=5in]{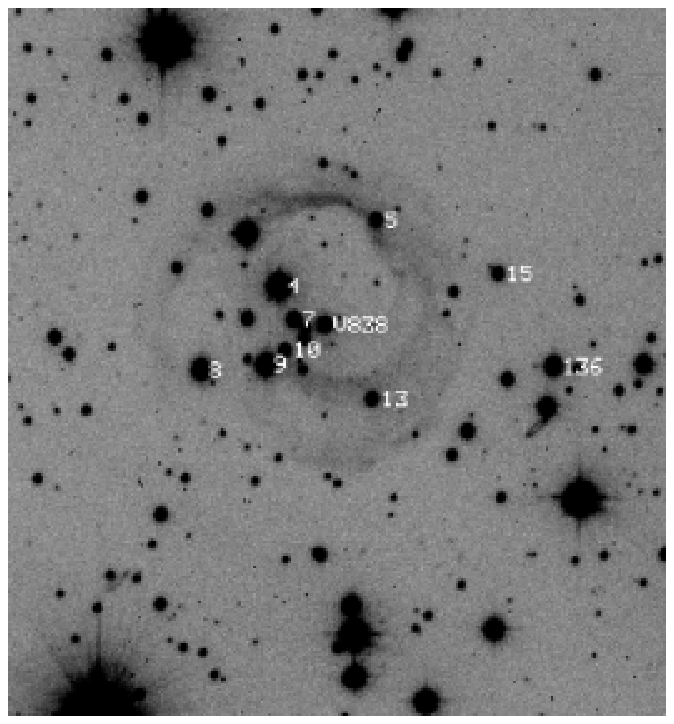}
\end{center}
\figcaption{Finding chart for stars in the vicinity of V838~Mon, prepared from
$B$-band exposures totalling 2000~s obtained with the SMARTS 1.3-m telescope on
2003 December 4. The image is $3\farcm6$ wide and has north at the top and east
on the left. The stars numbered 7, 8, and 9 are the newly discovered B-type
stars discussed in this paper. The other numbered stars have proven to be
unrelated foreground stars, based on our spectroscopic observations with the
SMARTS 1.5-m telescope.}
\end{figure}


\begin{figure}
\begin{center}
\includegraphics[width=6in]{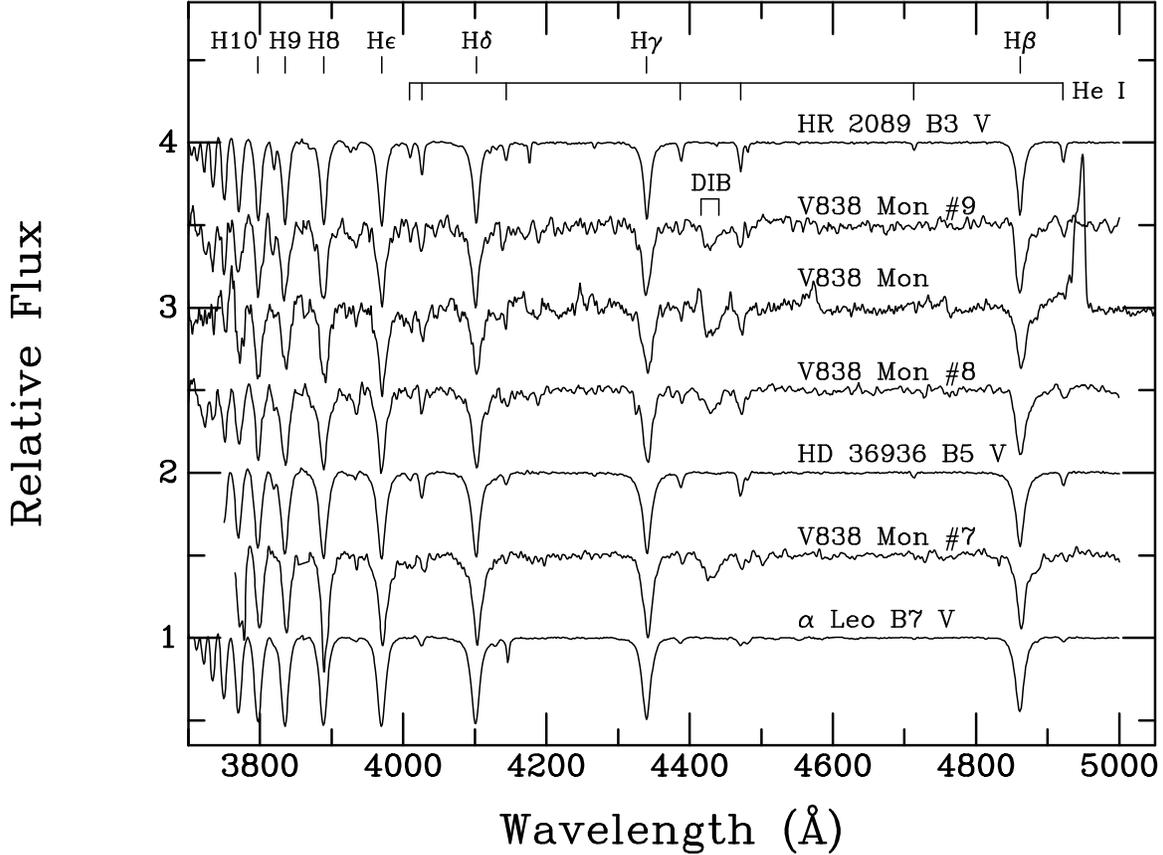}
\end{center}
\figcaption{SMARTS 1.5-m spectra of three B-type stars belonging to the young
cluster in the vicinity of V838~Mon, along with spectra of three classification
standards and V838~Mon itself (during early 2003). All spectra have been
normalized to a continuum level of 1.0, the tick marks on the $y$ axis are
separated by 0.5 continuum flux units, and the spectra have been offset by
constant successive amounts. A three-point boxcar smoothing has been applied to
the four faint stars. Stars 9, 8, and 7 are classified B3~V, B4~V, and B6~V,
respectively, by direct comparison with the classification standards, as
described in the text. We also classify the unresolved companion of V838~Mon
itself as B3~V\null. Note the strong diffuse interstellar band (DIB) at
4428~\AA\ in the four reddened stars belonging to the V838~Mon cluster.}
\end{figure}

\begin{figure}
\begin{center}
\includegraphics[width=6in]{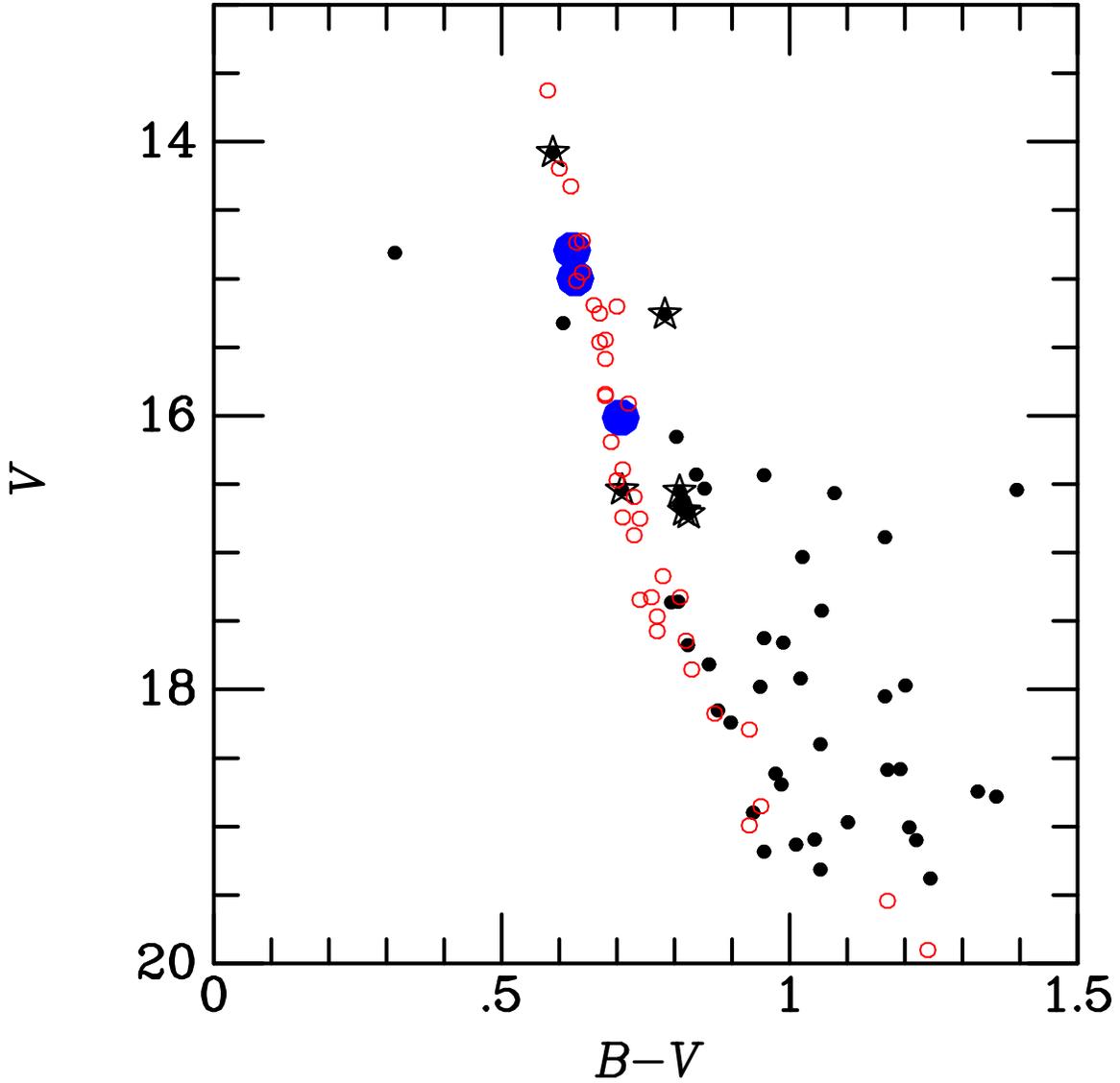}
\end{center}
\figcaption{Color-magnitude diagram for the three B stars belonging to the
V838~Mon cluster ({\it large blue filled circles}). Also plotted are all stars
within a $90''$ radius of V838~Mon ({\it small black filled circles}) for which
we do not have spectra, except for six spectroscopically confirmed foreground
field stars ({\it black star symbols}). The {\it open red circles\/} plot
photometry for the template zero-age main sequence of the open cluster NGC~2362
(Johnson \& Morgan 1953; Perry 1973), adjusted to a reddening of $E(B-V)=0.85$
and a distance of 6.2~kpc; field stars and binaries have been omitted.}
\end{figure}

\clearpage

\begin{deluxetable}{lccllccc}
\tablecaption{B-type Stars Near V838 Mon (J2000 Coordinates)}
\tablewidth{0pt}
\tablehead{
\colhead{Star No.} &
\colhead{WBM2003 No.} &
\colhead{M05 No.} &
\colhead{$\alpha$} &
\colhead{$\delta$} &
\colhead{Sp.~Type} &
\colhead{$V$} &
\colhead{$B-V$}
}
\startdata
7 & 2 & 197 & 07:04:05.48 & $-$03:50:49.3 & B6 V & 16.02 & 0.71 \\
8 & 5 & 159 & 07:04:07.47 & $-$03:51:06.1 & B4 V & 15.00 & 0.63 \\
9 & 3 & 110 & 07:04:06.05 & $-$03:51:03.6 & B3 V & 14.79 & 0.62 \\
\enddata
\end{deluxetable}

\begin{deluxetable}{lcccccccc}
\tablecaption{Spectroscopic Parallax Calculations}
\tablewidth{0pt}
\tablehead{
\colhead{Star No.} &
\colhead{Sp.~Type} &
\colhead{$V$} &
\colhead{$B-V$} &
\colhead{$(B-V)_0$} &
\colhead{$E(B-V)$} &
\colhead{$V_0$} &
\colhead{$M_V$} &
\colhead{$(m-M)_0$}
}
\startdata
7 & B6 V & 16.02 & 0.71 & $-0.15$  &  0.86  &  13.42 &  $-0.9$ & 14.32 \\
8 & B4 V & 15.00 & 0.63 & $-0.19$  &  0.82  &  12.40 &  $-1.4$ & 13.80 \\
9 & B3 V & 14.79 & 0.62 & $-0.205$ &  \underline{0.825} &  12.19 &  $-1.6$ &
  \underline{13.79} \\	 
  &   &  &      &         & \llap{Mean: } 0.84 & & & 14.0 \\
\enddata
\end{deluxetable}

\end{document}